\documentclass[11pt]{article}
\usepackage[T1]{fontenc}
\usepackage{deauthor}
\usepackage{times}
\usepackage{graphicx}

\newcommand{\mysceal}{MySc{\'e}al}

\graphicspath{{authorname/}}

\begin{document}

\title{Lifelogging As An Extreme Form of Personal Information Management - What Lessons To Learn}

\author{Ly-Duyen Tran$^1$~~~~~~Cathal Gurrin$^{1,2}$~~~~~~Alan F. Smeaton$^{1,3}$\thanks{Contact: \texttt{Alan.Smeaton@DCU.ie}}\\ \\
$^1$School of Computing\\
$^2$ADAPT Centre\\
$^3$Insight Centre for Data Analytics\\
Dublin City University, 
Glasnevin, Dublin 9\\
Ireland.}

\maketitle

\begin{abstract}
Personal data includes the digital footprints that we leave behind as part of our everyday activities, both online and offline in the real world. It includes data we collect ourselves, such as from wearables, as well as the data collected by others about our online behaviour and activities.  Sometimes we are able to use the personal data we ourselves collect, in order to examine some parts of our lives but for the most part, our personal data is leveraged by third parties including internet companies, for services like targeted advertising and recommendations. Lifelogging is a form of extreme personal data gathering and in this article we present an overview of the tools used to manage access to lifelogs as demonstrated at the most recent of the annual Lifelog Search Challenge benchmarking workshops. Here, experimental systems are showcased in live, real time information seeking tasks by real users. This overview of these systems' capabilities show the range of possibilities for accessing our own personal data which may, in time, become more easily available as consumer-level services.
\end{abstract}

\section{What is Personal Information Management~?}

Personal information or personal data is the digital form of the footprints of the everyday things that we do in our daily lives. It includes data that we can generate ourselves for example from wearable devices, or information that we ourselves create directly like our emails. In the case of the former, data from wearable devices can include physiological indicators like heart rate, respiration rate, heart rate variability or galvanic skin response levels, which are indicators of stress.  It can also include our movement data from accelerometers on as well as location data from GPS or other location-tracking technologies.  When such raw data is analysed then we can infer things like step counts, distances walked, run or biked, metrics for our sleep like duration and sleep quality and other health and wellness indicators. The purpose of capturing such personal data about ourselves is usually self-monitoring and self-tracking, typically for our own health awareness.    

While self-tracking is now a consumer-level rather than a specialist activity with hundreds of thousands of apps for uses like counting calories or fitness tracking  \cite{lupton2017self} much of what we call our personal data is actually recorded by others and not directly by ourselves. This includes the logs and records of our online activities like logs of our web browsing, social media interactions, media consumption and our communications and interactions with others.   Even though this data has been gathered and is being used by others, it is personal data that belongs to us.
Personal data gathered by third parties has value because of the insights it can reveal about us  and especially when it is aggregated and combined with the personal data of others \cite{birch2021data}. This form of personal data is exploited almost exclusively  by others, albeit supposedly for our benefit through targeted advertising and recommendations, and not directly by us.

Personal information management refers to  tools and platforms that allow us to control our personal data by allowing us to gather, store, update, analyse, interpret and sometimes to share it, no matter who has gathered it.
It is awkward or at best inconvenient for us to access or even to use our own personal data for our own personal purposes whether it is data we have gathered, or has been gathered by third parties.
While the log files of our interactions with search, browsing, social media, communications and media consumption services, were originally intended for helping to  debug and refine those systems and while we can request such logs from internet services it is not made easy for us to do so. Even for our use of systems in our own enterprises and organisations, like access logs to virtual learning environments in Colleges and Universities, this  typically requires downloading CSV files of personal data and then processing them ourselves, which is off-putting to most.  Access to our personal data from wearable devices is also typically supported through offering CSV or equivalent files as downloads. 

It is clear that our personal data can be a rich source of insights into us, into our well-being, our habits and behaviours and most importantly, into changes into those behaviours. Unfortunately we are able to query and to interrogate the individual and unconnected repositories for our personal data only on a per-device or per-source basis whereas the real benefits of such interrogations would be when the whole of our personal data, integrated across sources, could be analysed.  This poses the question of whether there is any work done that brings these sources together. 

In the later part of this article we will look at some very limited attempts to pool together our personal data from multiple sources but before that, in the next section, we will examine the current best-practice in the area of self-generated recording of everyday activities, the area known as lifelogging.

\section{Lifelogging: Extreme Personal Information Management}

In this section, we take a closer look as an extreme form of personal information gathering, which is called lifelogging~\cite{INR-033}. Lifelogging involves the comprehensive and continuous collection of data about one's daily activities and experiences, gathered from a wide array of data sources. Wearable devices, including smartwatches and fitness trackers, can be used to capture health metrics, such as heart rate, sleep quality, and activity levels. GPS sensors can be used to track one's movements and location. The most prominent data source for lifelogging is the use of wearable cameras to capture point-of-view images and videos of one's daily activities. These data sources provide a holistic representation of one's life and can be used for a variety of applications, such as health monitoring, memory prosthetic, and behavioural analysis.

The applications of lifelogging are widespread  and have been explored in many domains, such as supporting human memory recollection~\cite{barnard2011exploring,berry2007use,harvey2016remembering}, supporting large-scale epidemiological studies in healthcare~\cite{signal2017children}, monitoring the lifestyle of individuals~\cite{wilson2018use,nguyen2016recognition}, behaviour analytics~\cite{everson2019can}, and diet/obesity analytics~\cite{zhou2019use}. However, the vast volume and immense complexity of lifelog archives presents a challenge for users to navigate and analyse these archives in order to identify relevant information. As such, there is a growing interest in developing lifelog retrieval systems that effectively leverage lifelog data to meet the diverse needs of users.

Retrieval systems for lifelog data have been a popular research topic for several years. Managing the large amount of data collected by lifeloggers is crucial in order to maximise benefits from the data. In the MyLifeBits system, introduced more than 20 years ago by Gordon Bell \cite{10.1145/357489.357513}, then state-of-the-art techniques from database search and traditional information retrieval (IR) were employed to index and provide access to lifelog data through a desktop interface. However, as the volume of lifelog data increases, the need for more efficient and effective retrieval systems  becomes more apparent and the limitations of MyLifeBits soon became apparent. 

The first retrieval system  designed for large archives of lifelog data was proposed by Doherty {\em et al.}~\cite{doherty2012experiences}, moving the time/date \textit{browsing} approach of lifelog systems at the time to a \textit{search} approach. The system employed event segmentation, event annotation and multi-axes search, which are the `who', `what', `when', and `where' axes of retrieval. However, without a large user base, it was difficult to define search use-cases in order to evaluate and improve lifelog retrieval systems. 

In order to explore the possibilities and limitations of question-answering on the unique characteristics of timeline information such as that gathered in lifelogs, the authors of \cite{tan-etal-2023-timelineqa} have created and released TimelineQA. This is a generator that produces synthetic lifelogs for imaginary people with different personae including age, gender, education and family status. It uses a fine-tuned LLM to produce the actual content on which the effectiveness of question-answering can be investigated.  The synthetic lifelogs are composed of timelines of ordinary life events each of which have start times, end times and possibly locations, and they may also have multimodal features like photos. The TimelineQA benchmark is successful in that it is allowing a new form of interaction with lifelogs -- question-answering -- to be explored in a systematic and repeatable way.

\begin{figure}
    \centering
    \includegraphics[width=0.8\textwidth]{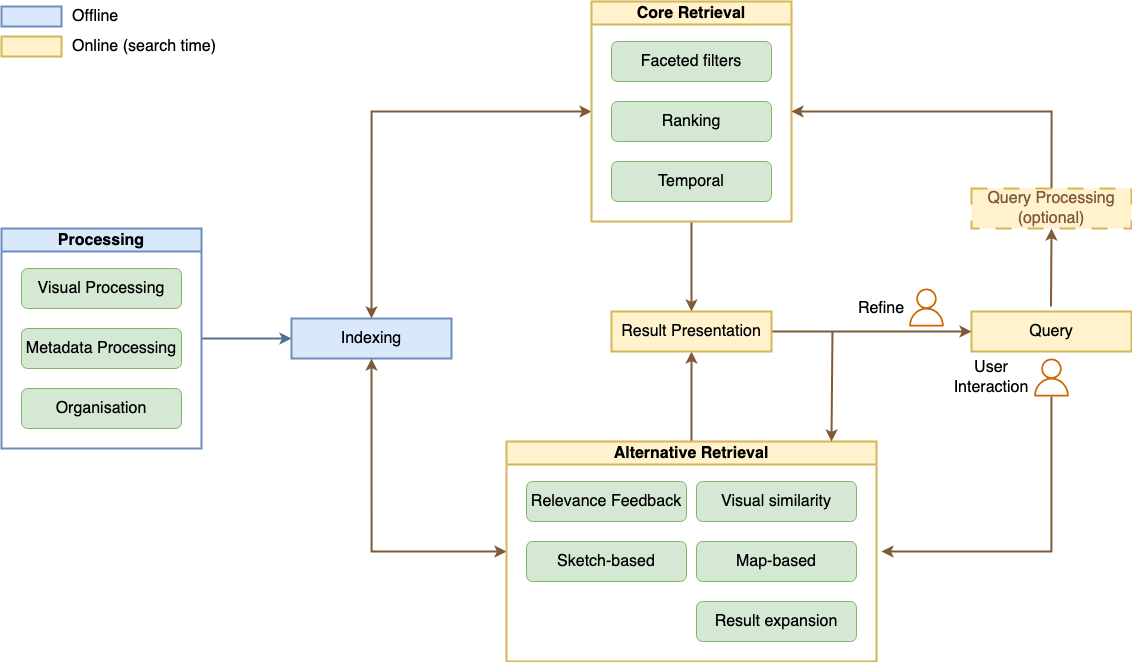}
    \caption{General pipeline of lifelog retrieval systems in the annual Lifelog Search Challenge (LSC).}
    \label{fig:lsc_pipeline}
\end{figure}

Several other benchmarking activities for lifelog retrieval systems have been organised which evaluate and compare the retrieval effectiveness of participating systems on real rather than synthetic lifelogs. The notable and best example of this is the annual Lifelog Search Challenge (LSC)~\cite{tran2023comparing}. The LSC provides a platform for evaluating  state-of-the-art systems for  managing lifelogs and  attracts a large number of participants internationally. In the LSC, different systems compete with each other in a live/virtual environment, where  participants are given a set of information needs and queries and a limited time to find the relevant and matching lifelog moments, in real time. Generally, most systems participating in the LSC follow the flow of data processing, indexing, and then retrieval, as shown in Figure~\ref{fig:lsc_pipeline}. Interactivity and user experience are  important aspects of the lifelog retrieval systems, as their affordances are known to directly influence the effectiveness of the retrieval process \cite{liu2019task}. We now discuss some of the techniques and components involved in lifelog retrieval with  example systems from the LSC. 

\subsection{Processing}
Lifelog data is typically collected from  sources including wearable cameras, wearable sensors, and mobile devices. The data is then processed to extract relevant information before being indexed and stored in a database. The processing stage is crucial as it determines the quality of the data and the effectiveness of the retrieval process.

\subsubsection{Visual Processing}
The most challenging aspect of processing is visual data, where different techniques are employed. We categorise these techniques into three groups: low-level features, concepts, captions, and embeddings.

Low-level features, such as colours, textures, and shapes, are commonly used in lifelog systems. These are classic features that have been used in the field of image processing for decades and are still used in lifelog systems. Their most popular application is the removal of low-quality images as seen in systems such as \mysceal~\cite{emysceal2022} (or its variant, E-\mysceal) and Memento~\cite{alam2023memento}. It is believed that low-quality images are less likely to be relevant to users, and thus, removing them can improve retrieval performance.

Due to  advances in computer vision algorithms, many lifelog systems utilise deep convolutional neural networks (CNNs) such as AlexNet~\cite{krizhevsky2012imagenet}, VGG~\cite{vgg}, GoogLeNet~\cite{szegedy2015going}, and ResNet~\cite{resnet}, trained on various dataset such as ImageNet~\cite{deng2009imagenet}, OpenImages~\cite{kuznetsova2020open}, Places365~\cite{zhou2014learning}, and Visual Genome~\cite{krishna_visual_2016}. These models are used to extract semantic lifelog `concepts', which are then indexed and used for retrieval. Optical character recognition (OCR) is also used to extract text from images, which proved to be useful in scenarios like identifying brand names and street names~\cite{tran2023comparing} during the real time search. This approach, often referred to as concept-based retrieval, is widely adopted due to its ease of implementation and the availability of pre-trained models.

Captioning, the generation of textual descriptions for images, is less commonly used in lifelog systems due to its inherent challenges. Generated captions are not always accurate, which limits its popularity. The most notable example of captioning in lifelog systems is vitrivr~\cite{spiess2023best}, which supports text-based search using generated captions. 

Embedding-based retrieval is an approach that has gained popularity in recent years. This involves mapping images into a vector space, where the similarity between images can be measured using cosine similarity, for example. This approach is widely used in cross-modal retrieval, where images are mapped into the same vector space as text queries. The most notable example of this approach is CLIP~\cite{radford2021learning}, which is a pre-trained model that maps images and text into a shared vector space. This approach replaces the need for concept extraction and captioning, as the similarity between images and text can be measured directly. Therefore, many lifelog systems have replaced their concept-based retrieval with embedding-based retrieval, as demonstrated by E-\mysceal~\cite{emysceal2022}, by adding embeddings to the existing concept-based retrieval and thus the name E-\mysceal. Other systems such as Memento~\cite{alam2023memento} and LifeSeeker~\cite{nguyen2023lifeseeker} have also adopted this approach. This \textit{embedding-based retrieval} approach allows a more user-friendly experience by allowing users to search for images using natural language queries and is shown to significantly improve  retrieval performance 

\subsubsection{Metadata Processing}
While metadata processing in lifelog systems is minimal due to standardised CSV formats provided by the LSC challenge organisers, it still plays a vital role in enhancing retrieval performance. The two most common metadata used in lifelog systems are timestamps and GPS coordinates. These help at retrieval time as users often remember aspects of the date or time or of the approximate location, of their personal data they are trying to find. Time information such as the time of the day, day of the week, month, and year are extracted from timestamps and used for filtering and scoring, as demonstrated by \mysceal~\cite{emysceal2022} and LifeSeeker~\cite{nguyen2023lifeseeker}. Clustering GPS coordinates for stay point detection, and inferring semantic locations are common techniques used to enhance the location metadata. For example, MyEachtra~\cite{tran2023myeachtra} employs a visual-aware stay point detection algorithm~\cite{tran2023vaisl} to detect stay points and utilises crowdsourced location data from FourSquare\footnote{https://developer.foursquare.com/docs/api-reference/} to infer semantic locations. Meanwhile, LifeSeeker~\cite{nguyen2023lifeseeker} enhances the location metadata by manually labelling images with `areas' within a location (such as kitchen, living room, and bedroom) to create a more fine-grained location hierarchy. This approach provides more `concepts' for users to filter and search for.

Biometric sensor data, surprisingly, is rarely used in lifelog systems due to noisy data and the lack of standardised formats. However, some systems have demonstrated the potential of using biometric data for retrieval by binning the data into categories~\cite{tran_lifelog_2018,alsina2018interactive,nguyen2019two}. Another type of metadata that was used in lifelog systems is music listening data. LifeSeeker uses music listening data to infer a user's mood, which is then used to filter and score the search results~\cite{nguyen2023lifeseeker}. However, due to the lack of LSC queries with the focus on these types of metadata, their effectiveness is yet to be evaluated.

\subsubsection{Organisation}
Due to the temporal nature of lifelog data, it is straightforward to organise lifelog data chronologically~\cite{zhou_baseline_2017}. Some of the LSC participating systems use a more sophisticated approach to organise the data, such as by using segmentation. Most systems that include segmentation use a simple approach by comparing the visual similarity between two consecutive images~\cite{emysceal2022}. MyEachtra~\cite{tran2023myeachtra} clusters the GPS coordinates to infer the semantic locations and then organise the data by the semantic locations.

LifeGraph~\cite{rossetto2023multi} offers a novel approach and organises the lifelog data by using a graph structure, with images as the centrepoint of the schema. The graph is constructed by linking the images with the metadata and detected concepts, then is extended with Wikidata~\cite{vrandevcic2014wikidata} and COEL (Classification of Everyday Living)~\cite{bruton2019classification}. However, the authors later acknowledged that COEL played an insignificant role in the query expansion and that Wikidata was a more useful source of concepts.

\subsection{Search and Navigation}
The core aspect distinguishing various lifelog retrieval systems is their retrieval mechanism. This  significantly influences how data is processed, indexed, and presented. 

Filters are a widely employed technique that has proven to be effective in lifelog systems. Faceted filters can be applied to various modalities and are often categorised by attributes such as time of day, day of the week, month, year, location, number of people, biometrics, and lifelog concepts as these are the aspects of personal data that a user can  recall in their information seeking \cite{10.1145/3025453.3025838}. Fixed phrases are typically used for filtering values, presented through drop-down lists, checkboxes, or sliders. Some systems, like \mysceal~\cite{emysceal2022}, allow users to enter free text, with autocomplete suggestions for available lifelog concepts. 

Query expansion is a technique used to assist users in formulating queries. It typically involves concept suggestion using various sources like WordNet, ConceptNet, and Thesaurus.com, and employing models like BERT~\cite{devlin2018bert} for concept similarity. \mysceal~\cite{emysceal2022} introduced a free-text query form with filter value extraction and query expansion, a method later adopted by LifeSeeker~\cite{nguyen2023lifeseeker}.

A ranking mechanism is essential for any retrieval system in order to sort outputs before presentaiton. Lifelog concepts can be used to score images based on the number of matched concepts~\cite{zhou_baseline_2017}. Extending TF-IDF term weighting using concepts extracted from computer vision models is a popular approach, as demonstrated by \mysceal~\cite{emysceal2022} and by LifeSeeker~\cite{nguyen2023lifeseeker}. TF-IDF is a common technique in information retrieval to score the relevance of a document to a query. It is calculated based on the term frequency (TF) and inverse document frequency (IDF) of the query terms. TF scores are not as useful as they are in the field of IR since the terms (concepts) are oftentimes not repeated in a document (image). Therefore, the TF scores are often replaced by the confidence scores of the concepts extracted from various computer vision models~\cite{dogariu_multimedia_nodate}. The area of the object (or its the bounding box) can also be exploited as in aTFIDF, proposed by \mysceal~\cite{emysceal2022}. 

As previously mentioned, cross-modal embedding models have offered a new way of measuring the similarity between images and text in lifelog retrieval systems. Cosine similarity between the search query and the images is directly used to rank the result. Some optimisation methods such as FAISS~\cite{johnson2019billion} or KNN search~\cite{emysceal2022} can be used to speed up the search. Although there is some effort of fine-tuning the embedding models on lifelog models~\cite{tran2022exploration}, large-scale pre-trained models are more robust and are often used directly or in a weighted ensemble in many systems, with Memento~\cite{alam2023memento} being the most notable example.

In addition, support for temporal search  is available in many LSC systems, enabling users to combine temporally related queries. For example, a user can search for `eating apple before watching TV'. Multiple temporally ordered queries can be executed conditionally, with results re-ranked accordingly based on the scores of each query.

\subsection{User Interactions}
Relevance feedback mechanisms are utilised in several lifelog systems to enhance retrieval performance. This technique, commonly employed in Information Retrieval, enables users to provide feedback on search results. Users can label images as relevant or irrelevant, prompting the system to refine the search iteratively. For example, Exquisitor~\cite{khan2021exquisitor} incorporates relevance feedback, where a user's feedback influences the training of classifiers to retrieve a more relevant set of images. Similar mechanisms are applied in systems like SOMHunter~\cite{lokovc2021enhanced}, where a user selects relevant images, or in Exclusion of concepts, allowing users to specify concepts they wish to exclude from the results.

Visual similarity plays a significant role in lifelog retrieval systems, often used in conjunction with other retrieval methods. It can be employed to arrange result lists as seen in SOMHunter~\cite{lokovc2021enhanced} and lifeExplore~\cite{schoeffmann2023lifexplore}; or to offer an alternative means for users to explore lifelog archives as seen in \mysceal~\cite{emysceal2022} and LifeSeeker~\cite{nguyen2023lifeseeker}. Visual similarity can be computed based on various factors, including low-level features such as colour histograms and SIFT, content-based features, or cross-modal embedding techniques.

Sketch-based search is a compelling method implemented in some lifelog systems originating from the video search community. Systems like lifeExplore~\cite{schoeffmann2023lifexplore} and vitrivr~\cite{spiess2023best} allow users to draw a sketch of the image they seek. The system then returns visually similar images to the sketch. This method proves useful when users cannot describe an image they wish to locate in words but is less popular due to users' limited ability to convey a target image accurately.

Map-based search is another frequently employed technique in lifelog systems. Users can draw a rectangle on a map to narrow down the search space, focusing only on moments that occurred within that area. Systems like lifeExplore~\cite{schoeffmann2023lifexplore}, \mysceal~\cite{emysceal2022}, and vitrivr~\cite{spiess2023best} offer map-based search features, often utilising libraries like Leaflet.

\subsection{Result Presentation Techniques}
Result presentation in lifelog retrieval systems has evolved beyond conventional grid views. Some systems have explored alternative ways of presenting results to enhance user experience. For instance, lifeExplore~\cite{schoeffmann2023lifexplore} and SOMHunter~\cite{lokovc2021enhanced} utilise Self-Organising Maps (SOMs) to arrange images in a 2D map, clustering them based on visual similarity. An autopilot navigation mode, introduced by lifeExplore in LSC'20, ensures every image in the target area is `visited' once. 

To mitigate visual clutter caused by similar images in lifelog data, event clustering is employed. Systems like \mysceal~\cite{emysceal2022} perform offline event segmentation and display the highest-ranked image from each event. In contrast, LifeSeeker~\cite{nguyen2023lifeseeker} dynamically clusters images during retrieval based on visual similarity and temporal proximity, presenting the top three images from each cluster. 

Addressing the temporal aspect of lifelogs, \mysceal~\cite{emysceal2022} proposes presenting results in triplets, showing the immediate previous and next events alongside the target event to provide more context. However, this approach is selectively used, specifically when temporal queries are specified, as adapted by E-\mysceal in LSC in 2022.

Memento~\cite{alam2023memento} offers a unique visualisation of results through distribution charts, allowing users to modify filters directly by interacting with the charts. Transitional graph-based visualisation, proposed by LifeSeeker~\cite{nguyen2023lifeseeker}, shows location transitions between images in the result list, which is particularly useful for transportation-related queries.

The ability to browse a lifelog archive chronologically, often referred to as the `timeline view', is crucial for users to explore their lifelog data. Many systems support this feature with varying levels of granularity and designs. Exquisitor~\cite{khan2021exquisitor}, for example, designed a temporal context view resembling a video player with lifelog images as thumbnails. Users can play the video to view images chronologically and navigate to specific times by selecting thumbnails below the video. Adjusting the spacing between thumbnails is also possible, which can be implemented using different methods such as scaling factors, step sliders, or hierarchical levels, as seen in systems like LifeSeeker~\cite{nguyen2023lifeseeker} and \mysceal~\cite{emysceal2022}.

\subsection{Summary}

In summary, lifelogging, as an extreme form of personal information management, offers a unique opportunity to gather and manage personal information. The annual lifelog search challenge event (LSC) serves as a remarkable example of how lifelog data can be harnessed and managed effectively. The techniques and components involved in the LSC represent the cutting edge of lifelog management, providing valuable insights into the future of personal data management. 

\section{Searching Personal Information}

In the previous section we have seen the potential for analysing and searching lifelogs as demonstrated in the annual Lifelog Search Challenge. All the systems participating in this interactive live benchmarking event are highly engineered and specialist and not designed for consumer-grade use, yet. 

In terms of our own personal data, we know that we can query and interrogate the individual, silo-ed and unconnected repositories which have generated that personal data which has come from individual sources, but is there anything in the personal information landscape that brings these together in some way to create an holistic overview~?
A recent  review of some personal data stores which can provide access to data from our online activities can be found at \cite{s23031477}. Some example systems included in that review include:

\begin{itemize}

\item {\bf Digi.me}: A platform that lets users collect their personal data from  online sources including social media, health, finance, and music, and to store it securely on their own devices or on the Digi.me cloud services. Users can then share their data with third parties on their own terms and get personalised insights and benefits from their data \cite{janssen2020personal};

\item {\bf Solid}: A project that aims to give users full ownership and control of their data by creating a decentralised web where they can store their data in personal online data stores called pods and link them to applications that respect their privacy and preferences \cite{jesus2020solid}.;

\item {\bf Meeco}: A platform that enables users to create a digital identity wallet where they can store and manage their personal data such as identity documents, credentials, preferences, and consent. Users can then use their wallet to access online services and share their data securely and selectively \cite{s23031477}.

\end{itemize}

\noindent 
While these examples of personal data stores may be useful for some use cases, they require the user her/himself to carry out the analysis of behaviour changes or patterns, and they are data stores rather than analysis platforms.  Thus they are not for non-technical users.
As an alternative to managing and analysing our own personal data we could invoke a third party to do the analysis, independently of the large internet companies. A platform to do this has been developed and demonstrated by Tuovinen \cite{tuovinen2022privacy} but that work is still early stage and not yet scaled up.

In the previous section we summarised the state-of-the-art in interactive lifelog search systems and showed that these systems have interactive question-answering capabilities which can lead to conversational retrieval around the topic of our own personal data. Outside the scope of the Lifelog search Challenge, the tools available to us for querying our own personal data do not yet exist for widespread use and those that are available no not have any forms of aggregation, or recommendation, or behavioural analysis, or self-reflection.

Within many areas of computing, the last year as seen a huge upsurge of interest in the use of large language models (LLMs) for applications in media creation, health and medicine, language processing, teaching and education, social interactions, science, and almost any domain where technology is used. While ChatGPT gets all the media coverage and popular focus, it is the fine-tuning of LLMs  on narrow applications which can be restricted in their scope, that holds the greatest long-term potential. The use of the Retrieval Augmented Generation (RAG) \cite{lewis2020retrieval} which allows an LLM to be fine-tuned on a limited training set could be applied to fine-tuning of an individual's personal data. This would allow the resulting system to leverage the advantages of LLMs including the capability for contextual understanding allowing a better understanding of a user's information need and search intent as well as the ability to simulate dialogue. Through careful prompt engineering it can also be used to unify across different sources of personal data and provide unified search, browsing and navigation across sources.  The first example we see of this is when in September 2023 Google announced that BARD AI, their conversational retrieval tool which uses Google's LLM, can integrate with a user’s emails, documents and other files as well as internet sources,  to give personalised answers to questions. It is thus an obvious next step that conversational retrieval supported by large language models, will be used by the major internet companies as they way for us to  access our personal data and that will certainly raise issues of data protection and privacy.

\section{Conclusions}

Personal information management involves the collection, storage, and control of digital data that represents our daily activities, including data generated by wearable devices, emails, and online interactions. This data can encompass various aspects of our lives, from physiological markers and indicators to our online behaviours. The purpose of gathering this personal data is primarily self-monitoring and self-tracking for health awareness. However, a significant portion of our personal data is collected and utilised by third parties, for targeted advertising and recommendations.

Personal information management tools and platforms should empower individuals to control our personal data, enabling us to gather, store, update, analyze, interpret, and sometimes share this data, regardless of its source. The development of such tools is essential because accessing and using personal data, even when it is our own data, is currently inconvenient and challenging due to the dispersed nature of our data across multiple sources. The paper presents an overview of an extreme form of personal data gathering called lifelogging and in particular how the annual lifelog search challenge highlights how innovative and useful searching through our lifelogs, our personal data, can be.

The article also highlights the potential of fine-tuning large language models (LLMs) for personalised data management, using techniques like Retrieval Augmented Generation (RAG). This approach could enable better contextual understanding, unified searching across various personal data sources, and the potential for conversational retrieval, thus revolutionising how individuals could access and manage their personal data.  With the rise of large language models and conversational retrieval, major internet companies are likely to play a significant role in providing access to our personal data, raising important questions regarding data protection and privacy in the future.

\bibliographystyle{abbrv}
\bibliography{bibfile}

\end{document}